\documentclass[prl,twocolumn,floatfix,superscriptaddress]{revtex4}
\usepackage{graphicx,amsfonts,amssymb,amsmath, hyperref,multirow,array}
\usepackage[applemac]{inputenc}

\newif\ifhyper
\hypertrue
\ifhyper
\hypersetup{
  citecolor = {green},
  colorlinks = {true}, 
  urlcolor = {blue} 
}
\fi

\newlength{\ldag}
\begin{document}

\title{Dynamical yield criterion for granular matter from first principles}

\author{O. Coquand} 
\email{oliver.coquand@umontpellier.fr}
\affiliation{Institut f\"ur Materialphysik im Weltraum, Deutsches Zentrum f\"ur Luft- und Raumfahrt (DLR), 51170 K\"oln, Germany}
\affiliation{Laboratoire Charles Coulomb (L2C), Universit\'e de Montpellier, CNRS, 34095 Montpellier, France}


\author{M. Sperl} 
\email{matthias.sperl@dlr.de}
\affiliation{Institut f\"ur Materialphysik im Weltraum, Deutsches Zentrum f\"ur Luft- und Raumfahrt (DLR), 51170 K\"oln, Germany}
\affiliation{Institut f\"ur Theoretische Physik, Universit\"at zu K\"oln, 50937 K\"oln, Germany}



\begin{abstract}
	We investigate, using a recently developed model of liquid state theory describing the rheology of dense granular flows, how a yield stress
	appears in granular matter at the yielding transition.
	Our model allows us to predict an analytical equation of the corresponding dynamical yield surface, which is compared to usual models of solid fracture.
	In particular, this yield surface interpolates between the typical failure behaviors of soft and hard materials.
	This work also underlines the central role played by the effective friction coefficient at the yielding transition.
\end{abstract}

\maketitle

\section{Introduction}

	Understanding the way solid objects break is a question relevant to various areas of physics.
	At a fundamental level, the determination of the precise mechanism at the origin of solid failure --- be it via elasto-plastic models
	\cite{Lin14,Lin15,Aguirre18,Liu18,Nicolas18,Ferrero19a},
	statistical methods inspired from glassy physics \cite{Wisitsorasak12,Nandi14},
	or modified elasticity theories \cite{Dasgupta12,Dasgupta13,Moshe15,DeGiuli18a,DeGiuli18c,DeGiuli20,Lemaitre21} ---
	as well the still debated relationship between the brittle and ductile modes of failure
	\cite{Fielding00,Alava06,Bouchbinder14,Rainone15,Jaiswal16,Parisi17,Urbani17,Popovic18,Ozawa18,Ozawa20,Goff20,Barlow20}
	are very active fields of research.
	But this topic is also ubiquitous in applied physics and engineering for the study of failure of rocks, soils, and other geomaterials
	\cite{Ancey07,Huang10,Mehranpour16,Radjai17,Zeng18,Fei20,Singh20,Wang20},
	concrete \cite{Francois08,Zhang10,Pan13,Jiang14}, or cellular materials \cite{Abrate08}.
	A particularly successful approach consists in determining the yield surface of the solid, a curve that allows to determine whether a solid
	in a given state of stress will yield.
	Many different constructions of such an object have been proposed \cite{Drucker52,Paul68,Lade75,Goddard84,Matsuoka85,Brunn02,
	Bigoni04,Matsuoka06,Labuz12,Lagioia16,Fleischmann20}
	but for amorphous solids, for which determining the yielding point from ab-initio methods used in crystals \cite{Clouet21} is not possible,
	no generally accepted construction of a yield criterion has been determined yet.

	In this paper, we propose a study of the yielding transition based not on a theory of the solid, but of the liquid state.
	More precisely, using a theory developed to describe the non-Newtonian features of granular liquid flows \cite{Kranz10,Kranz13,Kranz18,Kranz20,Coquand20f,
	Coquand20g,Coquand21}
	in the limit of very low shear rates, we describe how an internal state of stress develops into the liquid as its behavior becomes more and more solid-like.
	Such a determination of a so-called \textit{dynamical yield criterion} has previously been done for the study of colloidal suspensions close to the mode coupling glass transition
	\cite{Brader09}.
	Although there is no evidence suggesting a complete equivalence between the dynamical yield criterion determined when approaching the yielding transition from the
	liquid side and the yield criteria studied in triaxial tests when breaking solids, it is reasonable to assume that some properties are preserved across the transition.
	To that extent, our work brings an original insight on the yielding problem with an approach which, contrary to many failure models, is based on fundamental principles.

	The reduction of our liquid state theory to an analytically solvable model, along the lines presented in previous works \cite{Coquand20g,Coquand21} allows us
	to derive an analytical expression of the yield surface for granular materials.
	This surface turns out to display interesting properties, such as the existence a priori of two continuously related fracture modes (a soft and a hard one).
	Furthermore, the solvability of the model allows us to present a critical analysis of some widely used yield criteria.
	Finally, we show that a definition of the effective friction coefficient from the symmetries of the stress tensor allows us to build a quantity which
	behaves very smoothly across the transition to the solid state, thereby pointing out a potentially crucial quantity to understand how solid order builds up
	in complex liquids.


\section{Introduction to fracture}

	Our aim in this section is not to give a complete review of the theory of fracture in solids.
	We present some of the features of usual yield criteria which are relevant to the following \cite{foot1}.
	For sake of clarity, we restrict ourselves to the most usual yield criteria, many more refined ones being simple variations around those.

	The state of stress in a solid piece of material can be represented by a stress tensor $\sigma$, which can be further decomposed onto irreducible
	representations of the SO$(3)$ group into a diagonal part (spin 0 representation) and a traceless, deviatoric component (spin 2 representation):
	\begin{equation}
	\label{eqSig}
		\sigma = P \, \mathbb{I} + \sigma' \,,
	\end{equation}
	where $P=$Tr$(\sigma)/3$ is the pressure and Tr$(\sigma') = 0$.
	The same decomposition can be applied to the strain tensor $\epsilon$.
	In particular, for elastic materials, the deviatoric parts of both tensors are related by $\sigma' = 2G\epsilon'$, which defines the shear modulus $G$ of the solid.
	Since most of the following work concerns shear fracture, the other elastic moduli shall not be discussed.

	Another useful representation of $\sigma$ is given by its three eigenvalues --- or principal stresses --- $\sigma_1$, $\sigma_2$ and $\sigma_3$,
	which do not depend on the basis used to represent the stress tensor.
	One can then define three invariants in the following way:
	\begin{equation}
	\label{eqInv}
		\begin{split}
			& I_1 = \sigma_1 + \sigma_2 + \sigma_3 \\
			& I_2 = \sigma_1\sigma_2 + \sigma_2\sigma_3 + \sigma_3\sigma_1 \\
			& I_3 = \sigma_1\sigma_2\sigma_3 \,.
		\end{split}
	\end{equation}
	Similar definitions can be used to define the invariants $J_2$ and $J_3$ from $\sigma'$.

	Most of the yield criteria can be understood in terms of an effective friction coefficient.
	This coefficient can be defined in the following ways:\\
	\textbf{\underline{Definition 1:}} In a solid, the effective friction coefficient $\mu$ compares the strength of the spin 0 and the spin 2 components of the stress tensor.
	It can then be expressed as: $\mu = \sigma_0/P$, 
	where the shear stress is $\sigma_0 = \sigma:\epsilon'/|\epsilon'|$, "$:$" denotes a full tensor contraction, and the norm of the deviatoric strain
	is $|\epsilon'|^2 = 2\epsilon':\epsilon'$. \\
	\textbf{\underline{Definition 2:}} For two pieces of a broken solid to be able to glide onto one another, they have to overcome the solid friction between the two blocks,
	as defined by Coulomb's law.
	The effective friction coefficient $\mu$ of the solid is defined as the ratio of tangential to normal stress on a given plane that has to be overcome for the solid to
	break along that plane.
	Both definitions are related to one another, although not equivalent \cite{Fei20}.

	Let us suppose that a piece of solid breaks as soon as the elastic energy $E'_{el}$ accumulated due to the deviatoric strain exceeds a certain
	proportion of the isotropic part of the elastic energy $E_{el}^{Tr}\propto P$.
	The former energy can be expressed as $E'_{el}=\sigma':\epsilon'/2=J_2/4G$.
	The corresponding shear stress is $\sigma_0=\sqrt{J_2/2G}$.
	Finally, using the definition 1 of the effective friction coefficient, the material yields when $\mu$ exceeds a characteristic value of the material,
	$\mu_{DP}$, which defines the Drucker-Prager yield criterion \cite{Drucker52}.
	It can also be written in terms of invariants as $I_1^2/I_2=C_{DP}$, where $C_{DP}$ is a constant.

	The main advantage of definition 1 is that it only depends on the symmetries of the stress tensor.
	However, determining $\epsilon'$ in a given experimental situation can be challenging.
	As a result, many yield criteria are based on the second definition of $\mu$.
	In that case, the main difficulty is to determine the fracture plane on which the tangential and normal stresses must be compared.

	\begin{figure}
		\begin{center}
			\includegraphics[scale=0.6]{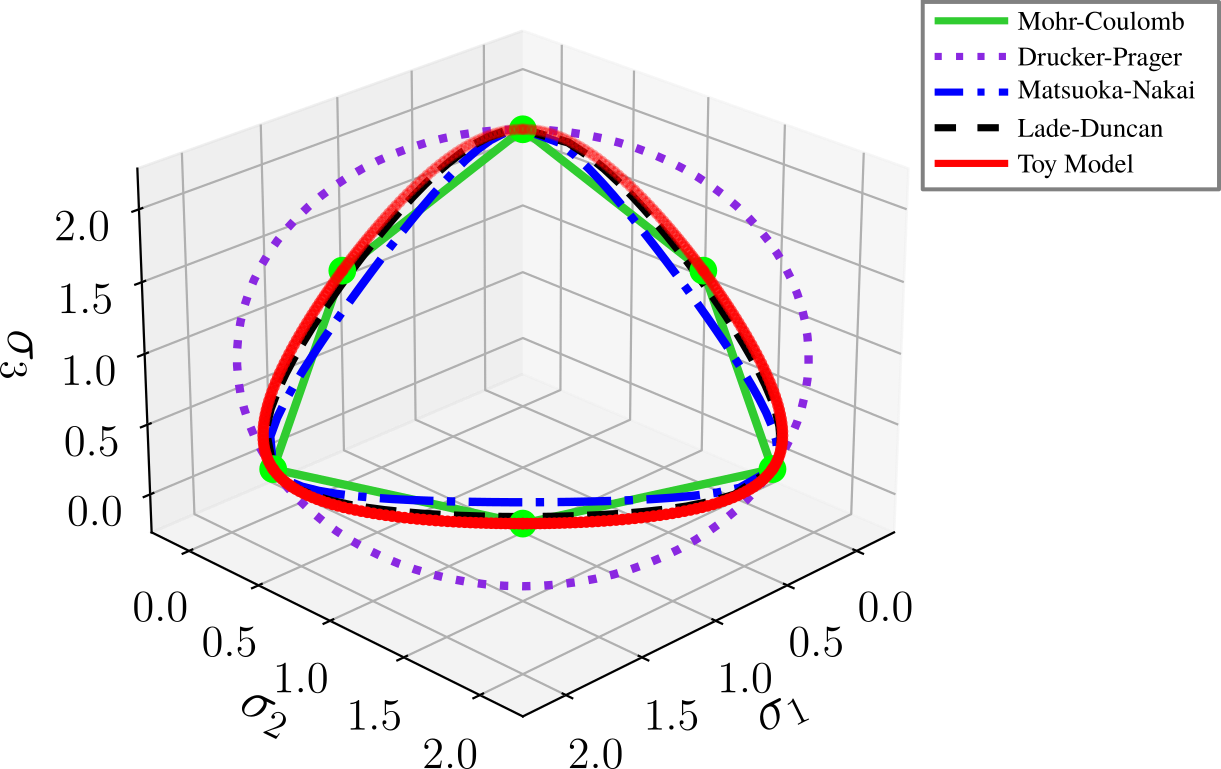}
		\end{center}
		\caption{Comparison of the different yield criteria presented in this paper.
		The big green dots are the points used to adjust the constants of the Mohr-Coulomb criterion (as well as their images
		by rotation of angle $2\pi/3$).
		The constants of the three other criteria are adjusted on the upper apex.}
		\label{FigYS}
	\end{figure}

	One of the simplest way to proceed is to assume that the fracture will take place in one of the principal planes.
	In that case, the problem reduces to three two-dimensional problems which are very easy to solve.
	The material yields as soon as $\mu\geqslant\mu_{MC}$ on one of the three principal planes, which defines the Mohr-Coulomb yield criterion.

	This approach can be refined further by approximating the fracture plane by the so-called \textit{spatially mobilized plane} \cite{Matsuoka85,Matsuoka06},
	which is a clever interpolation between the three principal planes.
	More precisely, it is defined as the unique plane which projections onto the three principal planes gives back the Mohr-Coulomb problem \cite{Matsuoka85}.
	By removing the possibility that the fracture plane evolves discontinuously from one principal plane into another, it provides a much smoother version
	of the Mohr-Coulomb yield surface (see Fig.~\ref{FigYS}).
	It terms of invariants, this can be written $C_{MN}=I_2I_1/I_3$, where $C_{MN}$ is a constant.

	Finally, another popular smooth version of the Mohr-Coulomb criterion is the Lade-Duncan yield criterion \cite{Lade75}.
	Its geometrical interpretation is not as straightforward as the previous ones, but it was recently showed \cite{Fleischmann20} that it can be understood as a correction to the
	Matsuoka-Nakai criterion taking into account subleading dependence in $J_2/P$.
	By cumulating the expressions of the Drucker-Prager and Matsuoka-Nakai constants, a new quantity $C_{LD}=C_{MN}C_{DP}=I_1^3/I_3$ can be constructed.
	The Lade-Duncan criterion can then be defined by requiring that $C_{LD}$, rather than $C_{MN}$ and $C_{DP}$ separately, is constant \cite{Fleischmann20}.
	The corresponding modification in terms of effective friction coefficient $\mu_{LD}$ is given in \cite{Fleischmann20}.

	A comparison of the different yield criteria is presented on Fig.~\ref{FigYS}, which presents a cut of the yield surface along one of the deviatoric planes
	(planes at constant pressure).
	Remarkably, the Drucker-Prager yield surface is much more isotropic than all the other ones.
	This can be related to the absence of dependence in $I_3$, which confers it a higher degree of symmetry.
	The more triangular shape of the three other criteria on the other hand can be related to the non zero value of the effective friction coefficient \cite{foot2},
	thereby illustrating the inequivalence between definition 1 and 2 given above ($\mu$ is obviously non zero too for the Drucker-Prager yield surface on Fig.~\ref{FigYS}).

	All in all, the most usual yield criteria used for studying solid fracture can be captured in terms of an effective friction coefficient $\mu$ which should not
	exceed a given value for solid order to be preserved.

\section{Results}

	\subsection{The yielding transition from the liquid state}

		The model we present in this paper is based on a simplification of the Granular
		Integration Through Transients (GITT) formalism.
		For more details about this model, the reader is referred to the more detailed previous
		publications on the subject \cite{Kranz10,Kranz13,Kranz18,Kranz20,Coquand20f,Coquand20g,Coquand21},
		as well as the appendix.
		This formalism is based on the so-called Integration Through Transients (ITT)
		formalism \cite{Fuchs02,Fuchs03,Fuchs09,Brader09}, which allows to compute
		statistical averages in a sheared fluid by relating them to averages computed in
		a quiescent state where no shear is applied.
		We can therefore decompose the stress tensor as: $\sigma= \sigma^{(0)} + \Delta\sigma$,
		where $\sigma^{(0)}=P_0\mathbb{I}$, because the quiescent fluid is not sheared, and
		$\Delta\sigma$ contains all the corrections depending on the shear rate $\dot\gamma$.
		The pressure $P_0$ can be computed for a fluid at equilibrium \cite{Coquand20f},
		but can also include an isotropic component imposed by the environment
		in many situations relevant to geophysics problems, and is thus left unspecified here.
		In our model, $P_0$ does not couple with the correction $\Delta\sigma$.

		The stress correction $\Delta\sigma$ can also be written in a form similar 
		to Eq.~(\ref{eqSig}):
		\begin{equation}
		\label{eqDSig}
			\Delta\sigma = \Delta P\mathbb{I} + \eta D\,,
		\end{equation}
		where $\Delta P$ is the correction to the pressure due to shear, $\eta$ is the viscosity
		of the liquid, and $D_{ij}=\kappa_{ij}+\kappa_{ji}$ is the symmetrized flow matrix,
		$\kappa=\nabla \textbf{v}$ being the velocity gradient.
		In this work, we restrict ourselves to incompressible flows for which Tr$(\kappa)=0$.

		The GITT equations allow to relate $\Delta\sigma$ to $\kappa$ via a viscosity
		matrix $\Lambda$ through $\Delta\sigma_{ab} = \Lambda_{abij}\kappa_{ij}$.
		In this formalism, the components of $\Lambda$ can be written explicitly as integrals
		over time and wave vectors of the density correlation function
		$\left<\rho_k(t)\rho_{-k}(0)\right>$ \cite{Coquand21}.
		Our toy model consists in neglecting the wave vector dependences which play
		a subleading role (details in appendix and \cite{Coquand20g,Coquand21}).
		In that case, $\Lambda$ reduces to integrals over time of products of the Finger tensor, which contains information
		about the deformation of the system, and the density correlation function (see the appendix for a detailed derivation).
		Then, the density correlation function is reduced to the following expression:
		\begin{equation}
		\label{eqTM}
			\left<\rho_k(t)\rho_{-k}(0)\right>\propto e^{-t/t_\Gamma}e^{-\dot\gamma^2 t^2/2\gamma_c^2}\,.
		\end{equation}
		The first factor accounts for the internal dynamics of the liquid, with a typical time scale $t_\Gamma$.
		As the behavior of the liquid becomes more and more solid-like, $t_\Gamma$ becomes very large, mostly due to the cage-effect:
		in dense liquids, particles' ability to move tends to be reduced by their neighbors.
		The second factor is a screening factor accounting for the effect of advection: the applied stress tends to force particles to move and facilitates the
		escape from the cages.
		This screening is characterized by a strain scale $\gamma_c$, which is a constant of the material describing its compliance to external stresses.
		Despite its simplicity, the reduction of the contribution to the stress tensor of the particle's dynamics by equations like Eq.~(\ref{eqTM})
		has proven to provide rather accurate constitutive equations for dense granular liquids \cite{Coquand20g,Coquand21}.

		The final element of our toy model is the sampling of possible flow geometries.
		First, we restrict ourselves to the basis in which $\sigma$ is diagonal, so that $\kappa$ is diagonal too.
		Then, we map the space of traceless diagonal matrices by the following two parameters family of reduced flow matrices:
		\begin{equation}
		\label{eqKa}
			\overline{\kappa} = \left(\begin{array}{ccc}
				A_2 & 0                                & 0 \\
				0   & -A_2\left(\frac{1-A_1}{2}\right) & 0 \\
				0   & 0                                & -A_2\left(\frac{1+A_1}{2}\right)
			\end{array}\right)\,,
		\end{equation}
		where $(A_1,A_2)$ define the flow geometry, and $\kappa = \overline{\kappa}\dot\gamma$.
		For this family of flows, and in the case of stationary flows, the toy model integrals giving the components of $\Lambda$ can be evaluated exactly.
		Finally, we can examine the limit $\dot\gamma\rightarrow0$, which allows two distinct behaviors: For low packing fraction systems, the ITT correction
		vanishes with $\dot\gamma$, the system remains a liquid; For denser systems, a non trivial yield stress develops in this limit, which signals the onset
		of solid-like behavior.
		The corrections to the eigenvalues of the stress tensor predicted by the toy model in the latter case can then be expressed as follows:
		\begin{equation}
		\label{eqYS}
			\begin{split}
				\Delta\sigma_1 &= \frac{S_0+S_1}{2}\sqrt{\pi}\big[\mathcal{F}(\overline{\kappa}_1\gamma_c) + \mathcal{F}(\overline{\kappa}_2\gamma_c) 
					+ \mathcal{F}(\overline{\kappa}_3\gamma_c)\big]  \\
					& \quad+ S_1\sqrt{\pi}\mathcal{F}(\overline{\kappa}_1\gamma_c) \\
				\Delta\sigma_2 &= \frac{S_0+S_1}{2}\sqrt{\pi}\big[\mathcal{F}(\overline{\kappa}_1\gamma_c) + \mathcal{F}(\overline{\kappa}_2\gamma_c) 
					+ \mathcal{F}(\overline{\kappa}_3\gamma_c)\big]  \\
					& \quad+ S_1\sqrt{\pi}\mathcal{F}(\overline{\kappa}_2\gamma_c) \\
				\Delta\sigma_3 &= \frac{S_0+S_1}{2}\sqrt{\pi}\big[\mathcal{F}(\overline{\kappa}_1\gamma_c) + \mathcal{F}(\overline{\kappa}_2\gamma_c) 
					+ \mathcal{F}(\overline{\kappa}_3\gamma_c)\big]  \\
					& \quad+ S_1\sqrt{\pi}\mathcal{F}(\overline{\kappa}_3\gamma_c) \,,
			\end{split}
		\end{equation}
		where $\mathcal{F}(x)=x\,e^{x^2}\,$erfc$(-x)$, and $S_0$ and $S_1$ are constants from the toy model giving the typical strength of the yield stress developing in
		the liquid (they have very little influence on the geometry of the yield surface).
		
		The yield surface equation Eq.~(\ref{eqYS}) is quite remarkable.
		Indeed, it was shown in \cite{Lagioia16} that all the usual yield criteria defined above can be expressed as roots of a polynomial equation of degree three.
		Here, to the contrary, the eigenvalues of the stress tensor, defined from the function $\mathcal{F}$ are highly non polynomial, which make Eq.~(\ref{eqYS}) quite
		unique to the best of our knowledge.
		As shown in the appendix, the Drucker-Prager criterion can be recovered from a $A_2\gamma_c\ll1$ expansion of Eq.(\ref{eqYS}), but the relation
		to the other usual yield criteria is more involved \cite{foot3}.

	\subsection{Geometry of the yield surface}

\begin{figure}
			\begin{center}
				\includegraphics[scale=0.7]{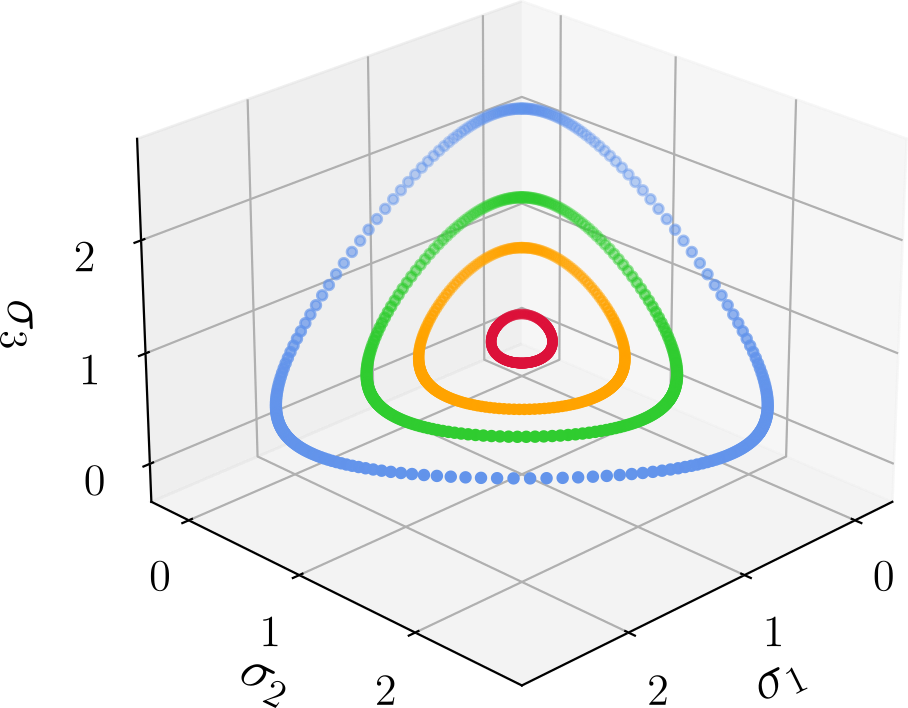}
			\end{center}
			\caption{Cuts of the toy model yield surface along the deviatoric plane for various values of $\Delta P$.
			The parameters are $S_0=S_1=1$, $\gamma_c=0.4$, $\Delta P = 0.05$ (red), $\Delta P = 0.5$ (orange), $\Delta P = 1$ (green),
			and $\Delta P = 2$ (blue).}
			\label{FigYSTM}
		\end{figure}

		On Fig.~\ref{FigYSTM}, we have represented the surface Eq.~(\ref{eqYS}) in various deviatoric planes, corresponding to various values of $I_1$.
		More precisely, fixing a pressure amounts to fixing the pressure correction $\Delta P$, which is our control parameter along the hydrostatic
		axis since $P_0$ does not couple to the shear corrections to the stress.
		All quantities are dimensionless, the global scale of the axis in physical units being fixed by the value of the parameters $S_0$ and $S_1$.

		The yield surface presents two qualitatively different behaviors from which we identify two fracture modes:
		For $\Delta P\simeq S_0,S_1$, the shape of the yield surface cuts is triangular, similar to that of Fig.~\ref{FigYS}, which corresponds to the
		typical shape observed in the failure of sedimentary rocks and soils captured by yield criteria from the family of Mohr-Coulomb/Matsuoka-Nakai/Lade-Duncan,
		we call this mode of fracture \textit{hard}; For $\Delta P\ll S_0, S_1$ on the other hand, the yield surface cuts become isotropic,
		and the yield surface has a Drucker-Prager like shape, we call this fracture mode the \textit{soft} mode.
		This can be used to classify material according to their preferred mode of fracture, softer materials being the ones for which the shear
		induced pressure component of the stress is small compared to the typical scale of stress involved.

		The soft fracture mode was already described in the yielding of soft colloids close to the mode coupling glass transition \cite{Brader09}.
		This picture fits well our current framework: close to the mode coupling glass transition, the shear correction to the pressure is rather mild
		compared to the other stress scales in the system.
		One of the main non trivial additional outcomes of our model is that this fracture mode is only observed under certain conditions (more precisely
		$A_2\gamma_c\ll1$, see the appendix for details), and can be continuously transformed into a hard type of fracture by changing the conditions of fracture.

		There are a number of reasons though why this transition between both fracture modes could not be observed experimentally:
		(i) While our model allows us to explore the variations of the yield surface with respect to any value of $\Delta P$, there is no guarantee
		that there exists an experimental protocol which allows to explore such a region for a given material.
		Indeed, $\Delta P$ is itself a function of the applied shear stress.
		(ii) Our model so far only includes shear failure, but it is expected that adding the possibilities of dilation and compression failure
		puts material dependent boundaries on the hydrostatic axis, which may prevent from exploring the full variation of the yield surface geometry.

		Finally, the analytical expression of the yield surface Eq.~(\ref{eqYS}) allows us to analyse the performance of other yield criteria in various deviatoric
		planes.
		The results are displayed on Figs.~\ref{FigCDP},\ref{FigCMN} and \ref{FigCLD} in the appendix.
		While unsurprisingly the Drucker-Prager criterion performs all the best that $\Delta P$ is small, there is no such monotonous behavior for the precision
		of the Matsuoka-Nakai and Lade-Duncan criteria.
		It can be noted, though, that those latter two perform all the best that the pressure $P_0$ is large.
	
	\subsection{The effective friction coefficient}

		We discussed in the first section how all usual yield criteria can be expressed in terms of an effective friction coefficient.
		Effective friction coefficients can also be defined for complex liquids.
		It is even a crucial quantity in the context of the study of granular liquids
		for which it has been shown that the dependence of $\mu$ on the shear rate
		is largely universal \cite{GDR04,DaCruz05,Jop06}, and provides a useful tool to relate the rheology of granular
		liquids and granular suspensions \cite{Cassar05,Boyer11,Houssais17,Guazzelli18,Coquand20g}.

		Since the definition of a fracture plane is not appropriate for liquids, $\mu$ is
		naturally defined from the symmetries of the stress tensor, like in definition 1 above.
		More precisely, the spin 0 component of the stress tensor of the liquid is the 
		total pressure $P=P_0 + \Delta P$, and its deviatoric component defines a shear
		stress $\sigma_0^l$ as $\sigma_0^l=\sigma:\kappa/|\kappa|$ \cite{Coquand21}, with $|\kappa|^2=D:D/2$.
		This defines an effective friction coefficient as $\mu=\sigma_0^l/P$.

		In a Newtonian liquid, the deviatoric stress is $\sigma'=\eta D$ (from Eq.~(\ref{eqDSig})).
		Because $\eta$ is a constant, this component of the stress vanishes in the limit $\dot\gamma\longrightarrow0$.
		For complex liquids however, $\sigma'=\eta(\dot\gamma)D$.
		Hence, provided that $\eta\sim1/\dot\gamma$ in the limit of low shear rates, this term survives and solid-like behavior builds up.
		All in all, analysing the deviatoric component of the stress tensor in the limit of low shear rates makes the connection
		between the liquid and solid definitions of the shear stress $\sigma_0$.
		However, both definitions of $\sigma_0$ are not necessarily equivalent to one another.
		Indeed, outside the regime of small strains, the deviatoric part of the deformation
		tensor is not proportional to the symmetrized velocity gradient $D$ \cite{Larson13}.

\begin{figure}
			\begin{center}
				\includegraphics[scale=0.5]{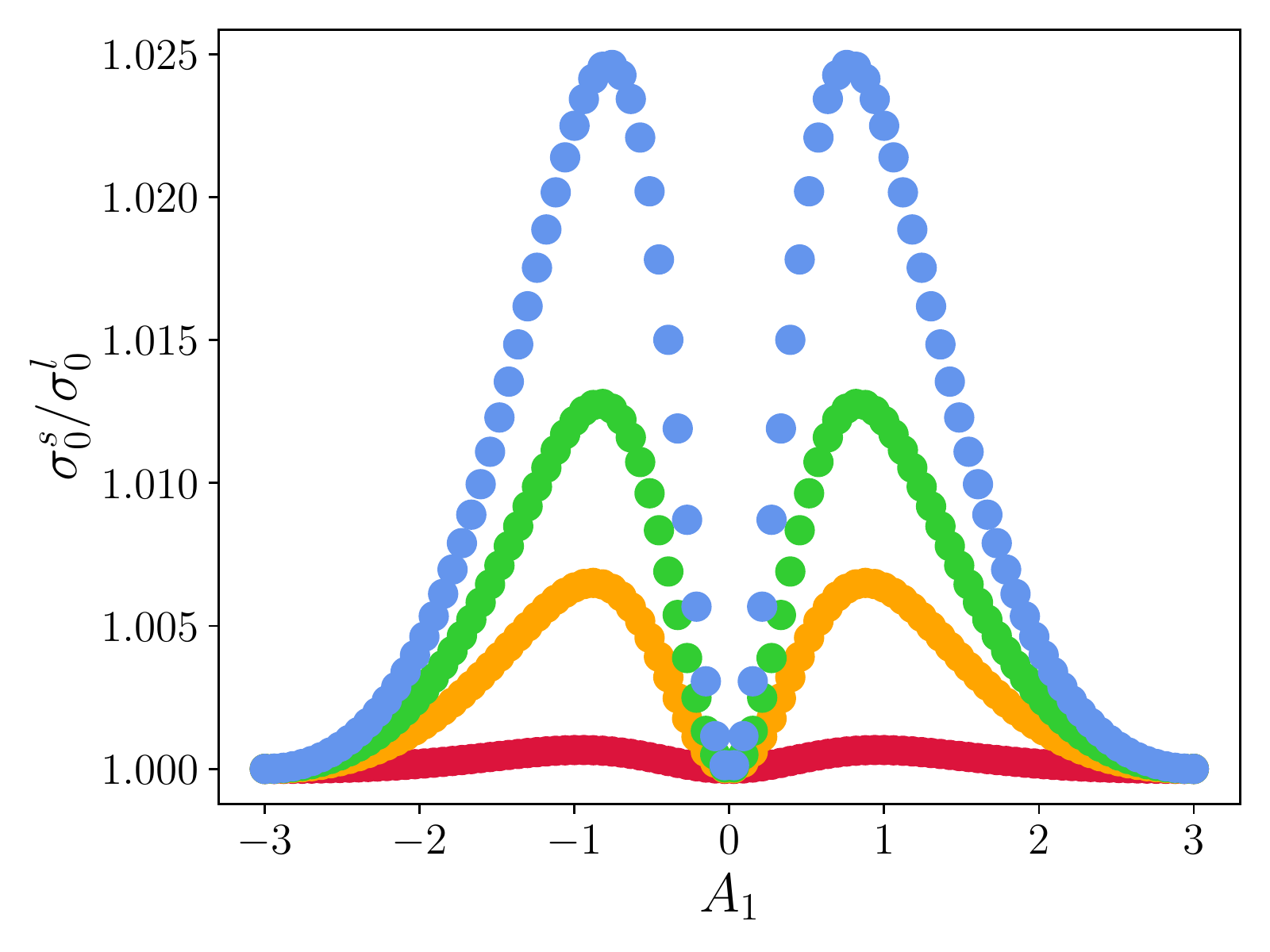}
			\end{center}
			\caption{Comparison of the definition of the shear stress from the solid state stress tensor $\sigma_0^s$ and from the liquid state stress
			tensor $\sigma_0^l$.
			The data is represented as a function of $A_1$, which variation on the presented scale correspond to a sector of $2\pi/3$ Lode angle.
			The parameters are $S_0=S_1=1$, $\gamma_c=0.4$, $\Delta P = 0.05$ (red), $\Delta P = 0.5$ (orange), $\Delta P = 1$ (green),
			and $\Delta P = 2$ (blue).}
			\label{FigRS0}
		\end{figure}

		Within our model, it is possible to identify from the stress equations (\ref{eqYS}) the contribution of the strain tensor (see the details in the appendix),
		and thus to get access to its deviatoric component.
		Hence, we can compare $\sigma_0^l$, computed from the velocity gradient, to $\sigma_0^s=\sigma:\epsilon'/|\epsilon'|$, computed from the 
		solid-like expression of the stress tensor Eq.~(\ref{eqSig}).
		The results are displayed on Fig.~\ref{FigRS0} for various values of $\Delta P$ \cite{foot4}.
		We can see that, although some variation is indeed present, both definitions yield very compatible numerical values.
		Consequently, $\mu$ defined from the ratio of the spin 2 and spin 0 component of the stress tensor behaves very smoothly across the
		liquid-solid transition, and therefore appears to be a particularly interesting quantity to study the onset of solid order in freezing liquids under shear.

		Finally, our model allows us to perform a more in-depth analysis of the behavior of $\mu$ in a given deviatoric plane (data in the appendix).
		First, $\mu$ is not constant along those planes, although relative variations are quite small.
		Provided that our model still holds for solid fracture, this means that a way to improve the existing yield criteria could be to allow for a such a deviation
		to the constant value, a feature which is present in none of the solid yield criteria presented in the first section of the paper.
		Then, the relative variation of $\mu$ with the Lode angle are all the bigger that the fracture mode becomes harder.
		Given that, as shown above, the hard fracture mode corresponds to cases very well described by criteria based on definition 2 of the effective friction
		coefficient, this raises the question of a possible transition between the two definitions as the isotropic stress $\Delta P$ caused by
		shear becomes more and more comparable to the intrinsic stress scale of the material fixed by $S_0$ and $S_1$: the definition of $\mu$ from the
		liquid stress tensor (\ref{eqDSig}) accounts for the build up of solid-like behavior as the liquid's behavior gets less and less Newtonian,
		then for the soft type of solid fracture mode, the approximation of $\mu$, set by definition 1, by a constant is quite accurate, but deteriorates as the
		fracture mode becomes harder, where, given the good performance of yield criteria based on definition 2, a constant $\mu$ based on definition 2
		could still be observed.
		Furthermore, since the value of $\mu$ on the fracture plane is determined by the value of the yield criterion constant \cite{Fleischmann20},
		we can also study the variations of $\mu$ from definition 2 within our toy model.
		Notably, there seems to be a regime where $P_0\simeq 1$ (in units of $S_0$, $S_1$) where the relative variation of $\mu_{MN}$ and $\mu_{LD}$
		is suppressed at best.
		However, since the value of $P_0$, playing a central role in that case, is not prescribed by our model, quantitative comparison to the relative variations
		with definition 1 is not possible.

\section{Conclusion}

	All in all, we have built an analytically solvable toy model that allows us to predict the shape of the dynamical yield surface of soft materials and granular
	matter.
	Importantly, our yield criterion is based on fundamental equations of liquid state theory rather than some phenomenological rule.
	The study of this model allowed us in particular to highlight the role of the effective friction coefficient $\mu$, that appears to be a particularly
	interesting quantity to relate the properties of the liquid close to the yielding transition, and those of the corresponding solid state.
	It also predicted a non trivial shape for the yield surface that smoothly interpolates between known shapes for soft and hard materials.

	The picture emerging from this study raises a number of questions:
	(i) How is the yield surface shape modified by the addition of the compression and dilation fracture modes ?
	This question can be answered by extending our toy model to compressible flows, in which case the liquid-like stress tensor provides both a shear and a bulk
	modulus in the limit $\dot\gamma\rightarrow0$.
	(ii) Is there a meaningful transition between definition 1 and 2 of the effective friction coefficient as the solid becomes harder ?
	Would an extension of our model to account for shear banding highlight some mechanisms of this transition ?
	(iii) What is the influence of the dependence of $\mu$ on the Lode angle, and thus on the stress geometry on the definition of the jamming transition,
	where $\mu$ is supposed to reach a fixed value through a power law evolution controlled by a critical exponent ?

\section*{Acknowledgements}

	This work was funded by the Deutscher Akademischer Austauschdienst (DAAD).
	We warmly thank Th. Voigtmann for stimulating discussions and helpful suggestions.
	We thank O. Pouliquen for enlightening discussions.

\appendix

\section{GITT equations and toy models}

	In this appendix, we quickly review the foundations of the toy-model used in the main paper.

	\subsection{Dynamics}

		The dynamics of the particles is accounted for through the evolution of the dynamical structure factor $\Phi_q(t)=\left<\rho_q(t)\rho_{-q}(0)\right>/S_q$, where 
		$\rho_q$ is the density operator in Fourier space, and $S_q=\left<\rho_q(0)\rho_{-q}(0)\right>$ is the static structure factor.
		Its evolution with time is given by a mode coupling equation of motion, which has the following schematic structure:
		\begin{equation}
		\label{eqMCT}
			\begin{split}
				\ddot{\Phi}_{q}(t) &+ \nu_{q}\dot{\Phi}_{q}(t)+ \Omega_{q}^2 \Phi_{q}(t) 
				+ \Omega_{q}^2 \int_0^t \!\!d\tau\,m_q(t,\tau)\dot{\Phi}_{q}(\tau) =  0 \,,
			\end{split}
		\end{equation}
		where $\nu_q$ and $\Omega_q$ are characteristic frequencies, and $m_q$ is called the memory kernel.
		The explicit expression of those terms is not needed for the following argument.
		The qualitative behavior of the solutions to Eq.~(\ref{eqMCT}) is as follows: When the memory term is small enough, Eq.~(\ref{eqMCT}) reduces to a linear second
		order differential equation, whose solutions have a decaying exponential envelope; When the memory term is large, typical solutions saturate to a finite
		plateau value at large time, signaling the onset of solid behavior.

		There are no analytical expression that captures well the global expression of the solutions to Eq.~(\ref{eqMCT}), hence building analytically solvable
		models require to do some approximations.
		A finer way to capture the behavior of $\Phi_q(t)$ is to use the Vineyard approximation, which decomposes it in a product of static structure factor and
		self correlation function, and then use an exponential ansatz for the latter term:
		\begin{equation}
		\label{eqVin}
			\Phi_q(t)\simeq S_q\,e^{-q^2\left<\Delta r^2(t)\right>}\,,
		\end{equation}
		where $\left<\Delta r^2(t)\right>$ is the mean squared displacement.
		It is not difficult to check that usual evolutions of $\left<\Delta r^2(t)\right>$ yields the expected qualitative phenomenology: for simple liquids, the
		large time behavior of $\left<\Delta r^2(t)\right>$ is given by the law of diffusion $\left<\Delta r^2(t)\right>\sim t$, which corresponds to the
		case of negligible memory effects with an exponential envelope of the decay, whereas in a solid, particles acquire a well-defined mean position,
		$\left<\Delta r^2(t)\right>$ reaches a finite limit value, and $\Phi_q(t)$ evaluated from Eq.~(\ref{eqVin}) does indeed saturate.

		From this qualitative analysis, we draw two main conclusions: (i) the wave vector dependence is irrelevant to understand the time evolution of $\Phi_q(t)$,
		at least in a lowest order approximation --- notice that the same type of argument is used when using schematic mode coupling models, such as those
		used in \cite{Fuchs03,Brader09} --- and (ii) the whole complexity of Eq.~(\ref{eqMCT}) can be captured in a very crude way by a simple exponential ansatz
		$\Phi_q(t)\simeq \exp(-\Gamma t)$, where $\Gamma = 1/t_\Gamma$ is structural relaxation rate.
		Of course, we do not pretend to capture all the mode coupling physics by this very simple ansatz, but it already captures the most relevant physical phenomena
		at play (in that case the slow down of the dynamics due to the structural relaxations in dense systems), and with only minor modifications (namely the addition
		of a second step in the decay of $\Phi_q(t)$), it provides a convincing model of the rheology of dense granular flows \cite{Coquand20f,Coquand20g}, allowing
		for example to recover the universal $\mu(\mathcal{I})$ of granular rheology.

		Finally, when going to the case of a sheared complex fluid, the effect of the advection of particles by the external shear stress has to be taken into account.
		Within our approach, this is done by evaluating the dynamical structure factor $\Phi$ at the wave vector $q(t)$ advected by the shear flow, rather than simply $q$.
		The equation (\ref{eqVin}) then becomes:
		\begin{equation}
		\label{eqVin2}
			\Phi_{q(t)}(t)\simeq S_q\,e^{-q(t)^2\left<\Delta r^2(t)\right>}\,,
		\end{equation}
		where $\mathbf{q}(t) = F(t)\cdot\mathbf{q}$, and $F(t) = \exp(-\kappa t)$ is the deformation gradient \cite{Larson13}.
		For simple shear flows, $\kappa$ is nilpotent, and $F(t) = \mathbb{I} - \kappa t$ exactly, but this relationship is not valid anymore for more general flows.
		Since $q(t)$ is typically ever increasing, this provides a new channel of decay for $\Phi$: Even when the packing fraction is so dense that $\left<\Delta r^2(t)\right>$
		should saturate to a constant value, the $q(t)^2$ factor in the exponential guarantees that $\Phi$ decays to zero.
		Physically, this corresponds to the fact that the motion imposed by the advection of particles by the applied shear stress is strong enough to break the cages, so that
		the systems always flow on some time scale.

		Because our toy-model is so far independent of $q$, it is blind to the change of $q$ to $q(t)$.
		To account for advection, we therefore add a screening term that introduces the advection channel of decay.
		Guided by results in the case of simple shear flows \cite{Coquand20g}, and the useful properties of Gaussian functions, we define the screening factor as
		$\exp(-\dot\gamma^2\,t^2/2\gamma_c^2)$, where $\gamma_c$ is a typical strain scale.
		As a matter of fact, the precise form of the screening is not really important to grasp the main properties of the rheology, \cite{Brader09} for example use
		a Lorentzian profile with similar results.

		All in all, at the level of the dynamics, our toy-model can be summarized by $\Phi_{q(t)}(t)\simeq\Phi_{toy}(t)=\exp(-\Gamma t)\exp(-\dot\gamma^2 t^2/2\gamma_c^2)$.
		It has two parameters: the structural relaxation rate $\Gamma$, and the strain scale $\gamma_c$.

	\subsection{Link to the rheology}

		The dynamics evolution is then linked to the rheology by use of the Integration Through Transients (ITT) formula \cite{Fuchs02,Fuchs03,Fuchs09,Brader09,Kranz20}.
		Using this approach, the shear correction to the stress tensor is expressed as an integral over the time evolution of a fictitious reference state, where the fluid
		is not sheared:
		\begin{equation}
			\Delta\sigma_{\alpha\beta} = \frac{1}{2T}\int_0^{+\infty}dt\int_k\,\mathcal{V}_{k(-t)}^\sigma\Phi_{k(-t)}(t)^2\mathcal{W}^\sigma_{k,\alpha\beta}\,,
		\end{equation}
		where the mode coupling vertices $\mathcal{V}_k^\sigma=\sum\kappa_{\theta\omega}\mathcal{V}_{k,\theta\omega}^\sigma$ and $\mathcal{W}_{k,\alpha\beta}^\sigma$
		have been introduced, and $\int_k=\int d^3k/(2\pi)^3$.
		Without going into the details, the vertices can be expressed as:
		\begin{equation}
		\label{eqVW}
			\begin{split}
				& \mathcal{V}_{k,\alpha\beta}^\sigma = \hat{k}_\alpha\hat{k}_\beta\,k\Delta\sigma + \delta_{\alpha\beta} \sigma_\perp \\
				& \mathcal{W}_{k,\alpha\beta}^\sigma = \frac{1+\varepsilon}{2S_k^2}\big[
					\hat{k}_\alpha\hat{k}_\beta\,k\Delta\sigma + \delta_{\alpha\beta} \sigma_\perp\big] \,,
			\end{split}
		\end{equation}
		as a function of the following reduced scalars
		\begin{equation}
		\label{eqsDO}
			\begin{split}
				& \sigma_\perp = T\big[S_k - S_k^2\big] \\
				& \Delta\sigma = - T S_k' \,,
			\end{split}
		\end{equation}
		the restitution coefficient of the granular particles $\varepsilon$, $S_k'=d\,S_k/dk$, and the normalized wave vector components $\hat{k}_\alpha$.
		Note that the above formula can be applied to colloidal suspensions by taking the elastic limit $\varepsilon\rightarrow1$.

		Given that we defined above a procedure of approximation of the $\Phi_{k(-t)}^2(t)$ term, the next step is to reduce the tensorial structure of the mode
		coupling vertices.
		Factoring out the $\kappa_{\theta\omega}$ term embedded in the vertex $\mathcal{V}_k^\sigma$, our integral becomes a tensor of rank four corresponding
		to the viscosity tensor $\Lambda_{\alpha\beta\theta\omega}$.
		The vertex product is then developed:
		\begin{equation}
		\label{eqAB1}
			\begin{split}
				\mathcal{V}^\sigma_{k,\theta\omega}\mathcal{W}^\sigma_{k,\alpha\beta}\propto & \delta_{\alpha\beta}\delta_{\theta\omega}\sigma_{\perp}^2
				+ \delta_{\theta\omega}\hat{k}_\alpha\hat{k}_\beta k \sigma_\perp\Delta\sigma \\
				& + \delta_{\alpha\beta}\hat{k}_\theta(-t)\hat{k}_\omega(-t)k(-t) \sigma_\perp\Delta\sigma\\
				&+ \hat{k}_\alpha\hat{k}_\beta\hat{k}_\theta(-t)\hat{k}_\omega(-t) k k(-t)\Delta\sigma^2\,.
			\end{split}
		\end{equation}
		Then, we use the deformation gradient $F(t)$ introduced above to extract the remaining time dependence carried by the wave vectors by
		$\hat{\kappa}_\alpha(t) = F_{\alpha\nu}(t) \hat{k}_{\nu}$.
		The remaining time independent wave vector component being the only non isotropic terms, the spherical part of the $k$ integral can be performed.
		We use the following formula:
		\begin{equation}
		\label{eqAB2}
			\int_k \,k_ik_j\,f(k^2) = \frac{\delta_{ij}}{3}\int_k f(k^2)\,,
		\end{equation}
		and
		\begin{equation}
		\label{eqAB5}
			\int_k \,k_ik_jk_ak_b\,f(k^2) = \frac{X_{ijab}}{15}\int_k f(k^2)\,,
		\end{equation}
		where we defined the fully symmetrized product of kronecker symbols as:
		\begin{equation}
		\label{eqABX}
			X_{ijab} = \delta_{ij}\delta_{ab} + \delta_{ia}\delta_{jb}+ \delta_{ib}\delta_{ja}\,.
		\end{equation}
		Finally, $\Phi_{k(-t)}(t)$ is replaced by $\Phi_{toy}(t)$, and the remaining integral over the norm of $k$ reduces to a constant prefactor.

		After all the above steps are performed, the viscosity tensor can be decomposed along the three following terms:
		\begin{equation}
		\label{eqAB3}
			\begin{split}
				& \mathcal{B}^{comp}_{\alpha\beta\theta\omega} = \delta_{\alpha\beta}\delta_{\theta\omega}\int_0^{+\infty}dt\,\Phi_{toy}^2(t) \\
				& \mathcal{B}_{\alpha\beta\theta\omega}^0 = \delta_{\alpha\beta}\int_0^{+\infty}dt \,\Phi_{toy}^2(t)\big[\delta_{\theta\omega}
				+ F_{\nu\theta}(-t)F_{\nu\omega}(-t)\big]\\
				& \mathcal{B}_{\alpha\beta\theta\omega}^1 = X_{\alpha\beta\nu\iota}\int_0^{+\infty}dt \, \Phi_{toy}^2(t) F_{\nu\theta}(-t)F_{\iota\,\omega}(-t) \,.
			\end{split}
		\end{equation}
		In the above formula, $\mathcal{B}^{comp}$ comes from the $\sigma_\perp^2$ term. It never contributes to $\Delta\sigma$ for incompressible flows.
		$\mathcal{B}^0$ comes from the $\sigma_\perp\Delta\sigma$ term in Eq.~(\ref{eqAB1}), the constant prefactor defines $S_0$. It includes
		both the effect of restitution coefficient and the remnants of the $k$ integral. The first term in the bracket only contributes in the case of compressible flows.
		The second term is the Finger tensor, the inverse of the right Cauchy-Green deformation tensor.
		Lastly, $\mathcal{B}^1$ comes from the $\Delta\sigma^2$ term in Eq.~(\ref{eqAB1}) and its prefactor defines $S_1$.
		Thus, in the case of incompressible flows, two scaling constants --- $S_0$ and $S_1$ --- are needed, and all the contributions to $\Lambda$ have the form of a time
		integral of the product $\Phi_{toy}(t)$ and two deformation gradients, with varying types of tensor contractions.

		For the last step, let us place ourselves in the basis in which the flow matrix is diagonal.
		In that case, it is possible to show that the remaining contributions in Eq.~(\ref{eqAB3}) only involve one component of $\kappa$ at a time, say $\kappa_{jj}$.
		A typical term will hence have the following form:
		\begin{equation}
		\label{eqInt}
			\begin{split}
			\kappa_{jj} & \int_0^{+\infty}dt \,\exp\left(-2\Gamma t - \frac{\dot\gamma^2 t^2}{\gamma_c^2} + \kappa_{jj} t\right) \\
			&= \frac{\sqrt{\pi}}{2}\overline{\kappa}_{jj}\exp\big[\gamma_c^2 (\overline{\kappa}_{jj} +\Gamma/\dot\gamma)^2\big]
			\text{erfc}\big[\gamma_c(\overline{\kappa}_{jj} +\Gamma/\dot\gamma)\big]\,.
			\end{split}
		\end{equation}
		Finally, our study is restricted to the study of dense granular liquids and suspensions, close to the yielding transition, which Weissenberg numbers
		Wi are very large, hence $\Gamma\ll\dot\gamma$, so that Eq.~(\ref{eqYS}) is recovered.
		Note that in the mode coupling paradigm, such systems are characterized by $\Gamma=0$, because they would be solid if the system were not sheared.
		However, a weaker version of the argument is needed here: we only use the fact that Wi$\gg1$.
		This corresponds to taking a limit $\dot\gamma\rightarrow0$ under the constraint that $\Gamma\ll\dot\gamma$.

\section{The Drucker-Prager limit}

	Let us discuss the limit of small $\Delta P$.
	The value of $\Delta P$ can be deduced from Eq.~(\ref{eqYS}), and depends mainly on the behavior of the function $\mathcal{F}$.
	Hereafter are some of its basic properties:
	\begin{equation}
	\label{eqF}
		\begin{split}
			& \lim_{x\rightarrow+\infty}\mathcal{F}(x) = + \infty \\
			& \lim_{x\rightarrow-\infty}\mathcal{F}(x) = - \frac{1}{\sqrt{\pi}} \,.
		\end{split}
	\end{equation}
	Then, the typical size of the argument of $\mathcal{F}$ in Eq.~(\ref{eqYS}) is set by the combination $A_2\gamma_c$, $A_1$ fixing the relationship between the
	different components of $\kappa$ (see Eq.~(\ref{eqKa})).
	This combination contains both information of the strength of the shear flow, through $A_2$, and the material's response, through $\gamma_c$.
	Because of Eq.~(\ref{eqF}), and since Tr$(\kappa)=0$, which ensures that at least one of its components is positive, if $A_2\gamma_c\gg1$, $\Delta P$ is large.
	Consequently, the regime of small $\Delta P$ correspond to small or moderate values of $A_2\gamma_c$, which can be explored by an expansion in powers
	of $A_2\gamma_c$.

	The expansion of $\mathcal{F}$ around $x=0$ is as follows:
	\begin{equation}
	\label{eqFl}
		\mathcal{F}(x) = x + \frac{2}{\sqrt{\pi}}\, x^2 + O(x^3)\,.
	\end{equation}
	Because Tr$(\kappa) = 0$, the first order term in the expression of $\Delta P$ vanishes, leaving at leading order:
	\begin{equation}
	\label{eqDPtaylor}
		\Delta P = I_1 - P_0 = \frac{(3S_0 + 5 S_1)(3 + A_1^2)}{2}\,A_2^2\gamma_c^2 + O(A_2^3\gamma_c^3)\,.
	\end{equation}
	Since $I_1$ and $P_0$ are parameters of our model, the above relationship truncated at this order allows us to compute $A_1$ as a function of $A_2$, and the parameters
	of the model on a given deviatoric plane.

\begin{figure}
		\begin{center}
			\includegraphics[scale=0.7]{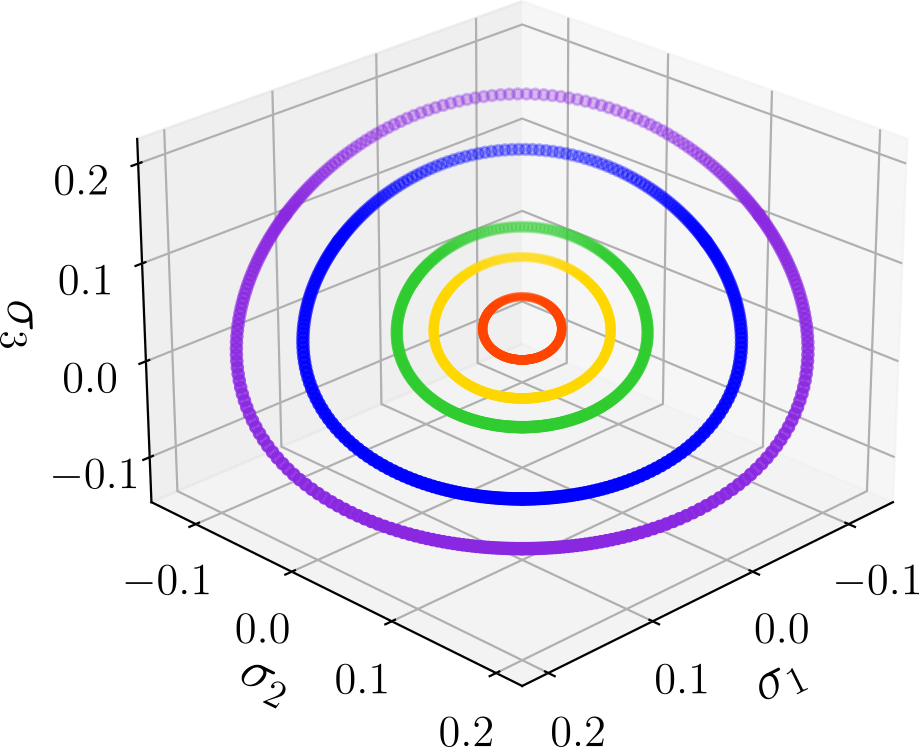}
		\end{center}
		\caption{Toy model yield surface for $\Delta P\ll S_0, S_1$.
		The parameters are $S_0=S_1=1$, $\gamma_c=0.4$, $\Delta P = 0.001$ (orange), $\Delta P = 0.005$ (yellow), $\Delta P = 0.01$ (green),
		$\Delta P = 0.03$ (blue), and $\Delta P = 0.05$ (violet).}
		\label{FigYSlP}
	\end{figure}

	The corresponding geometry of the yield surface can be deduced by studying the evolution of the stress tensor invariants.
	For example, applying the same expansion,
	\begin{equation}
	\label{eqJ2DP}
		J_2 = S_1^2\pi\,\frac{3 + A_1^2}{2}\,A_2^2\gamma_c^2 + O(A_2^3\gamma_c^3)\,,
	\end{equation}
	so that combining it with Eq.~(\ref{eqDPtaylor}) yields:
	\begin{equation}
		\frac{J_2}{I_1^2} = \frac{S_1^2\pi}{2}\frac{I_1 - P_0}{(3S_0 + 5S_1)I_1^2} + O(A_2\gamma_c) \,.
	\end{equation}
	This formula can be read as follows: (i) the cuts of the yield surface in a given deviatoric plane are circles and (ii) their radius is all the smaller that
	$\Delta P/(3S_0 + 5S_1)$ is small.
	It is the behavior of the Drucker-Prager yield surface.
	Such behavior is indeed observed with the yield surface of Eq.~(\ref{eqYS}), as can be seen on Fig.~\ref{FigYSlP}.
	As $A_2\gamma_c$ gets closer to 1, $\Delta P/(3S_0 + 5S_1)$ does too, and the yield surface cuts become less and less isotropic.

	Interestingly, the fact that the expression of the yield surface is known analytically also means that we can study the behavior of the constants associated with the
	other usual yield criteria in the same limit, namely the Drucker-Prager, Matsuoka-Nakai and Lade-Duncan ones.

	Let us first have a look at the Drucker-Prager constant $C_{DP} = I_1^2/I_2$.
	In order to have an idea of its typical values, it is instructive to have a look at the value of $J_2$ at the point of Lode angle equal to 0, which
	in our conventions corresponds to the uniaxial extension flow $\overline{\kappa}_{xx} = 1$, $\overline{\kappa}_{yy} = -1/2$, $\overline{\kappa}_{zz} = -1/2$.
	Indeed, this point is generally used to adjust the constants of various yield criteria, since it corresponds to the point of maximum $J_2$, or, said otherwise,
	the point furthest to the hydrostatic axis $\sigma_1 = \sigma_2 = \sigma_3$.
	In that point, $J_2$ can be expressed as a function of the Mohr-Coulomb friction angle $\phi_{MC}=$arctan$(\mu_{MC})$ as:
	\begin{equation}
		J_2 = \frac{4\,I_1^2\,\sin^2(\phi_{MC})}{3(3 - \sin(\phi_{MC}))^2} \,,
	\end{equation}
	so that, since $0\leqslant \mu_{MC} \leqslant + \infty$, $0\leqslant J_2/I_1^2 \leqslant 1/3$, and consequently, $3 \leqslant C_{DP}\leqslant + \infty$.
	From Eq.~(\ref{eqYS}), $C_{DP}$ can be expanded as:
	\begin{equation}
	\label{eqexpCDP}
		\frac{C_{DP}}{3} = 1 + 3 \mathcal{X}\left[1 + \frac{3}{4}(A_2\gamma_c)\frac{\pi - 8}{\sqrt{\pi}} \right] + O(A_2^3\gamma_c^3) \,,
	\end{equation}
	where we introduced the combination
	\begin{equation}
		\mathcal{X} = \frac{(I_1 - P_0)\pi^{3/2}S_1^2}{P_0^2\big[3S_0 + (3 + 4\sqrt{\pi})S_1\big]} \,,
	\end{equation}
	which appears in all the following expansion.
	$\mathcal{X}$ is all the smaller that $\Delta P$ is, and is constant in a given deviatoric plane.
	From Eq.~(\ref{eqexpCDP}), we can check that $C_{DP}$ does converge toward $1/3$ when $A_2\gamma_c$ is small, and becomes typically bigger than $1/3$ when
	$A_2\gamma_c$ grows.

	A similar analysis can be performed in the cases of the Matsuoka-Nakai and Lade-Duncan criteria.
	First, notice that the ratio of tangential and normal stresses on the spatially mobilized plane is given by:
	\begin{equation}
		\mu_{MN}^2 = \frac{C_{MN}}{9} - 1 \,,
	\end{equation}
	which holds independently of the precision of the Matsuoka-Nakai criterion ($\mu_{MN}$ is simply more or less constant).
	Since this ratio is positive by definition, $9\leqslant C_{MN}\leqslant +\infty$.
	The expansion of this coefficient in the limit $A_2\gamma_c\ll1$ is given by:
	\begin{equation}
	\label{eqexpCMN}
		\frac{C_{MN}}{9} = 1 + 6 \mathcal{X}\left[1 + \frac{3}{4}(A_2\gamma_c)\frac{P_0(\pi - 8) + 6\pi S_1}{P_0\sqrt{\pi}} \right] + O(A_2^3\gamma_c^3) \,,
	\end{equation}
	Again, this is consistent with the known limiting behavior of this coefficient.
	Note also that in the limit where the hydrostatic pressure dominates the intrinsic strength of the material $P_0\gg S_1$, the second term in the bracket becomes
	similar to that of Eq.~(\ref{eqexpCDP}).

	Finally, using $C_{LD} = C_{MN} C_{DP}$, $27 \leqslant C_{LD} \leqslant + \infty$ \cite{Fleischmann20}.
	The expansion of this coefficient in the limit $A_2\gamma_c\ll1$ is given by:
	\begin{equation}
	\label{eqexpCMN}
		\frac{C_{LD}}{27} = 1 + 9 \mathcal{X}\left[1 + \frac{3}{4}(A_2\gamma_c)\frac{P_0(\pi - 8) + 4\pi S_1}{P_0\sqrt{\pi}} \right] + O(A_2^3\gamma_c^3) \,,
	\end{equation}
	which is consistent with the boundaries.

\section{Analysis of the usual yield criteria}

\begin{figure}
		\begin{center}
			\includegraphics[scale=0.5]{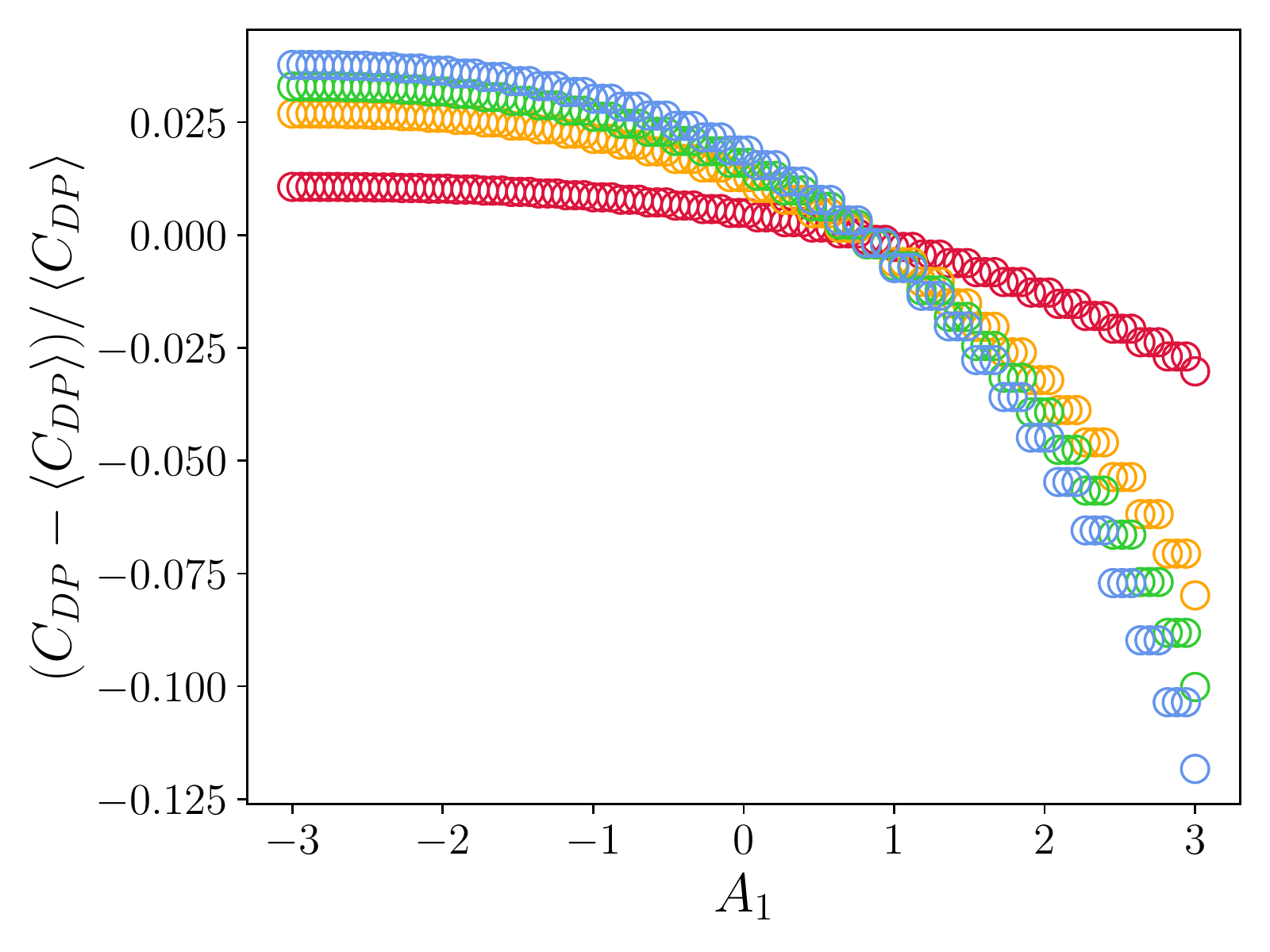}
		\end{center}
		\caption{Relative variation of the Drucker-Prager constant for various values of $\Delta P$.
		The parameters are $S_0=S_1=1$, $P_0=1$, $\gamma_c=0.4$, $\Delta P = 0.05$ (red), $\Delta P = 0.5$ (orange), $\Delta P = 1$ (green),
		and $\Delta P = 2$ (blue).}
		\label{FigCDP}
	\end{figure}

\begin{figure}
		\begin{center}
			\includegraphics[scale=0.5]{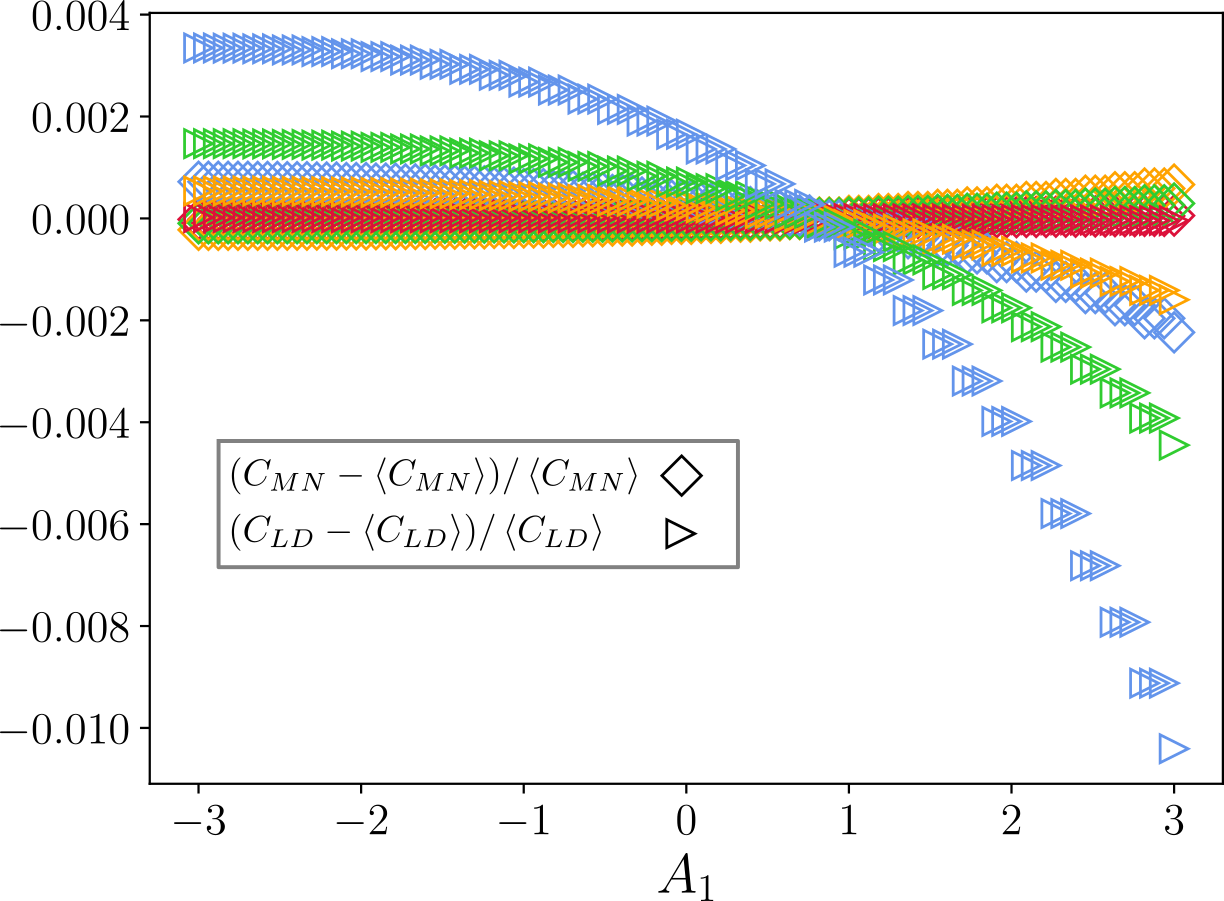}
		\end{center}
		\caption{Relative variation of the Matsuoka-Nakai (diamond) and Lade-Duncan (triangles) constants for various values of $\Delta P$.
		The parameters are $S_0=S_1=1$, $P_0=1$, $\gamma_c=0.4$, $\Delta P = 0.05$ (red), $\Delta P = 0.42$ (orange), $\Delta P = 1$ (green),
		and $\Delta P = 2$ (blue).}
		\label{FigCMN}
	\end{figure}

\begin{figure}
		\begin{center}
			\includegraphics[scale=0.5]{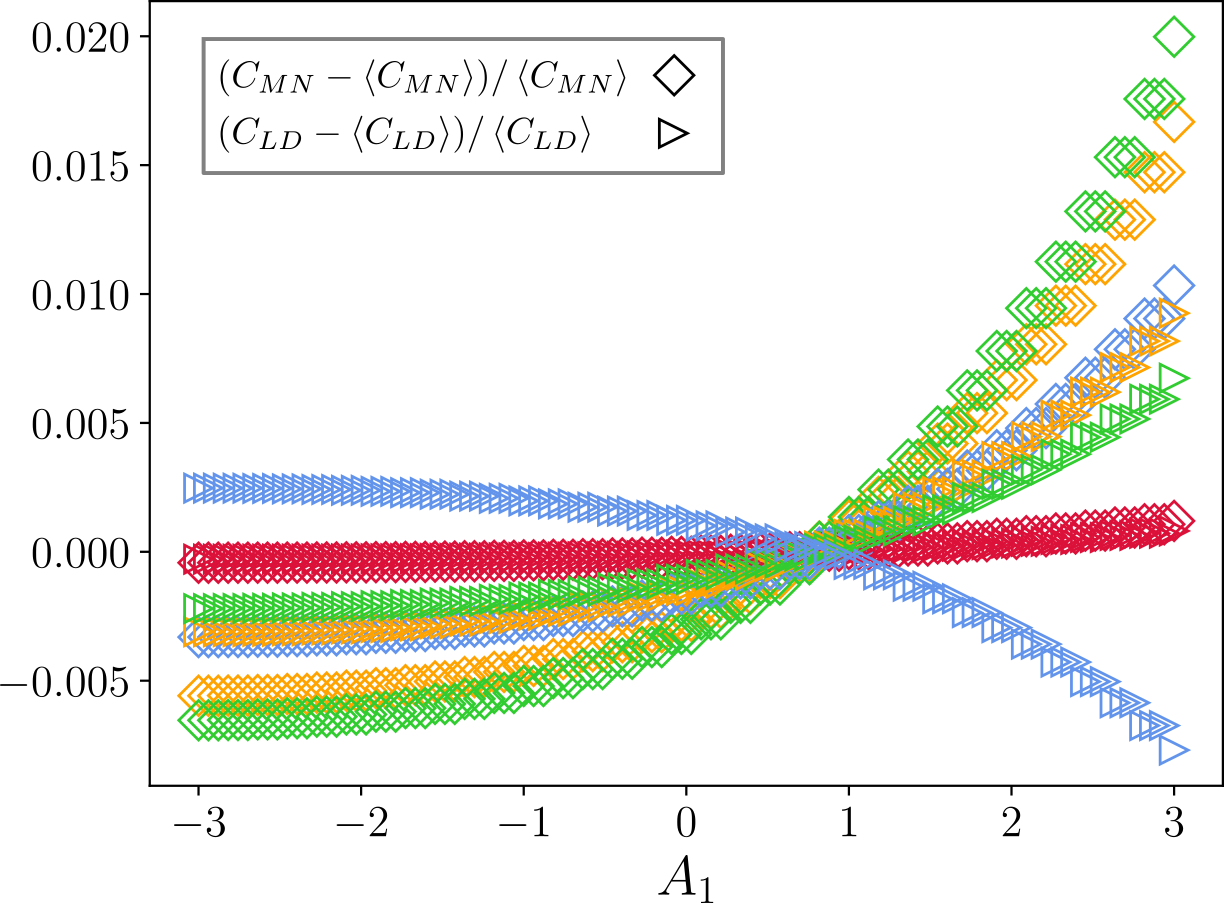}
		\end{center}
		\caption{Relative variation of the Matsuoka-Nakai (diamonds) and Lade-Duncan (triangles) constants for various values of $\Delta P$.
		The parameters are $S_0=S_1=1$, $P_0=0.5$, $\gamma_c=0.4$, $\Delta P = 0.05$ (red), $\Delta P = 0.5$ (orange), $\Delta P = 1$ (green),
		and $\Delta P = 2$ (blue).}
		\label{FigCLD}
	\end{figure}

	Most yield criteria can be represented by a certain combination of invariants of the stress tensor being constant.
	This type of assertion is easy to test within the realm of our toy model.
	For a given yield criteria of equation $f_x(I_1,I_2,I_3)=C_x$, where $C_x$ is a constant, we can evaluate $f_x$ on a given
	deviatoric plane (in most cases, models are tested for a given value of $I_1$), compute its average on the whole yield surface
	cut in that plane, $\left<f_x\right>$, and look at the relative variation $(f_x - \left<f_x\right>)/\left<f_x\right>$.
	Finally, in a given cut of the yield surface along a deviatoric plane, the data can be represented as a function of the Lode angle,
	or equivalently as a function of the parameter $A_1$ of our surface parametrisation.
	In this way, we get the figures \ref{FigCDP}, \ref{FigCMN} and \ref{FigCLD}.

	More precisely, the determination of $I_1$, $I_2$ and $I_3$ depend on $P_0$, which is fixed by the environment and thus not prescribed in our model.
	As can be expected, as $P_0$ becomes larger and larger compared to the values of the intrinsic stress scales $S_0$ and $S_1$, the approximation 
	of $C_{LD}$ and $C_{MN}$ by constant values is of better and better quality.

	\subsection{The effective friction coefficient}

\begin{figure}
		\begin{center}
			\includegraphics[scale=0.5]{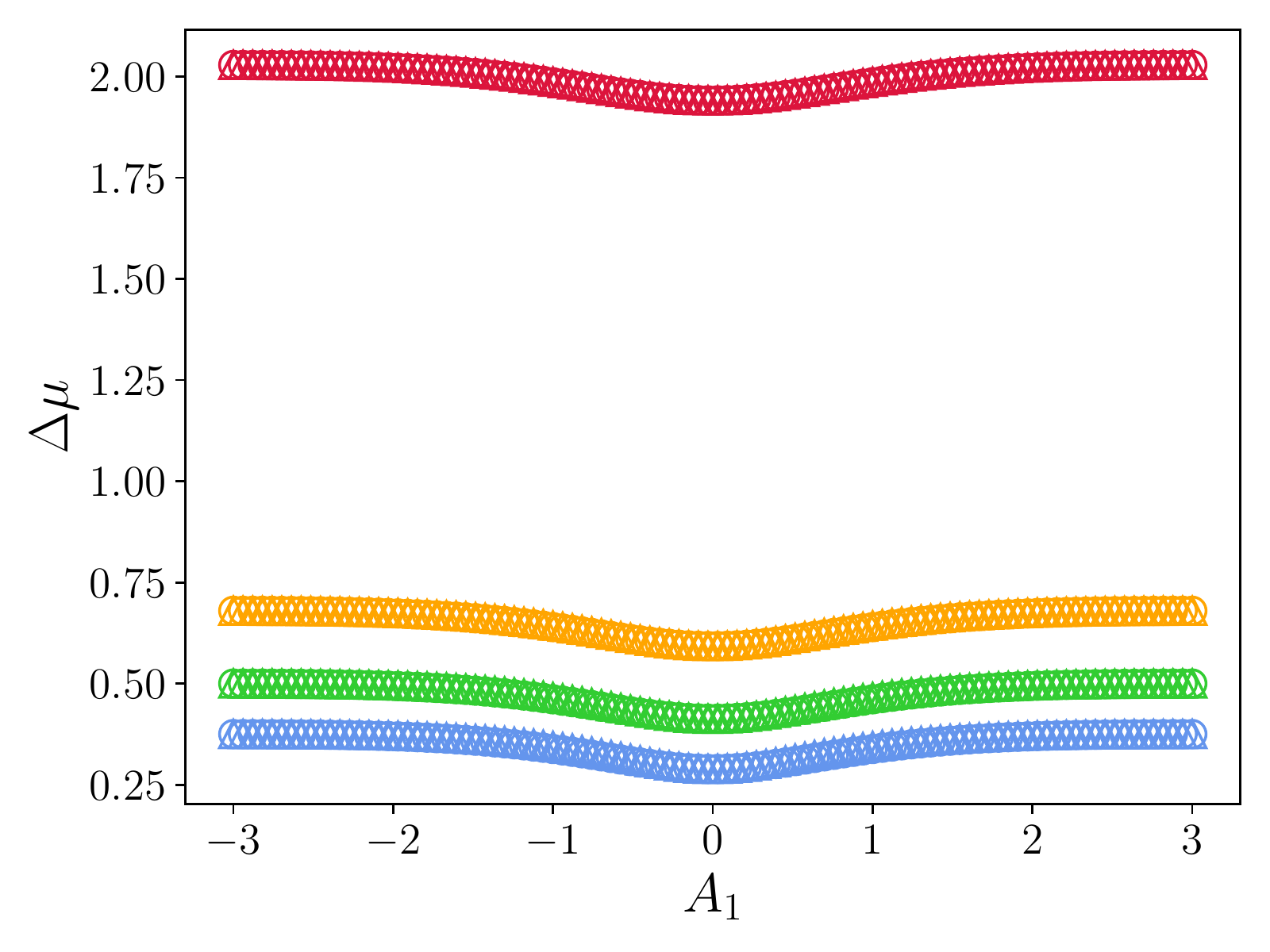}
		\end{center}
		\caption{Evolution of $\Delta\mu = \sigma_0/\Delta P$ as a function of $A_1$ for $\sigma_0=\sigma_0^s$ (triangles) and $\sigma_0=\sigma_0^l$ (circles).
		The parameters are $S_0=S_1=1$, $\gamma_c=0.4$, $\Delta P = 0.05$ (red), $\Delta P = 0.5$ (orange), $\Delta P = 1$ (green),
		and $\Delta P = 2$ (blue).}
		\label{FigDmu}
	\end{figure}

\begin{figure}
		\begin{center}
			\includegraphics[scale=0.5]{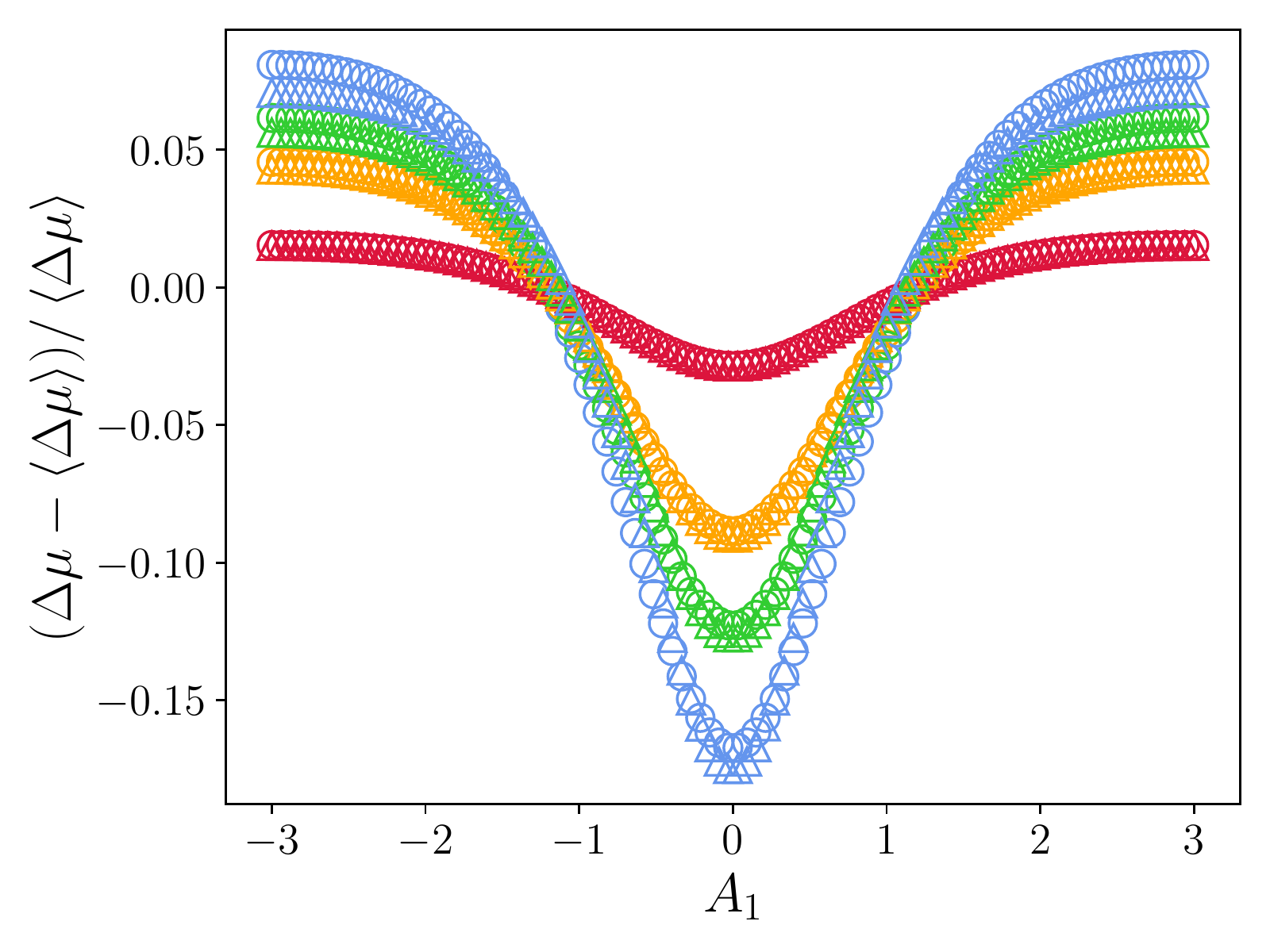}
		\end{center}
		\caption{Evolution of the relative variation of $\Delta\mu = \sigma_0/\Delta P$ as a function of $A_1$ for $\sigma_0=\sigma_0^s$ (triangles) and
		$\sigma_0=\sigma_0^l$ (circles).
		The parameters are $S_0=S_1=1$, $\gamma_c=0.4$, $\Delta P = 0.05$ (red), $\Delta P = 0.5$ (orange), $\Delta P = 1$ (green),
		and $\Delta P = 2$ (blue).}
		\label{FigDDmu}
	\end{figure}

		The effective friction coefficient is defined as a function of the total pressure $P=P_0 + \Delta P$. However, there is no prescription
		for $P_0$ in our formalism.
		The evolution of effective friction can still be observed through the study of the reduced friction $\Delta \mu=\sigma_0/\Delta P$.
		Its evolution as a function of the Lode angle (represented by $A_1$) for various values of $\Delta P$ is displayed on Fig.~\ref{FigDmu},
		while Fig.~\ref{FigDDmu} represents its relative variations.

		While looking at Fig.~\ref{FigDmu}, it should be kept in mind that what we represent is only the reduced friction coefficient, and not the total
		one.
		Therefore, relatively large values of $\Delta \mu$ must not be deemed too surprising, they are reduced by the total confining pressure $P_0$.

		We chose to represent $\Delta \mu$ for both definitions of the shear stress $\sigma_0$, from the solid and the liquid stress tensors respectively,
		in order to illustrate the very close proximity of the predictions in both frameworks.
		Indeed, this difference appears to be visible only for its relative variation, and only for the largest values of $\Delta P$ (in units of $S_0$, $S_1$).

\begin{figure}
		\begin{center}
			\includegraphics[scale=0.5]{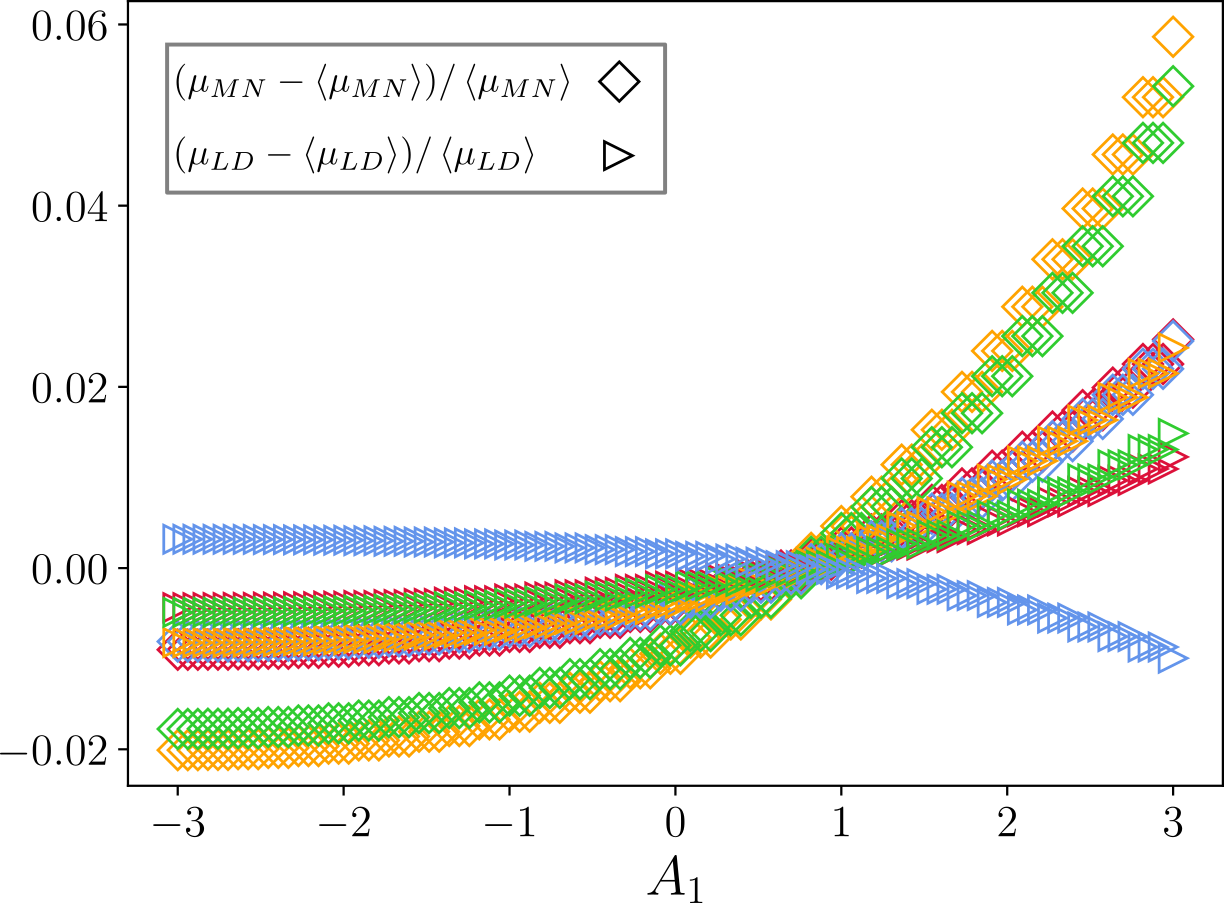}
		\end{center}
		\caption{Relative variation of the Matsuoka-Nakai (diamonds) and Lade-Duncan (triangles) effective friction coefficients for various values of $\Delta P$.
		The parameters are $S_0=S_1=1$, $P_0=0.5$, $\gamma_c=0.4$, $\Delta P = 0.05$ (red), $\Delta P = 0.5$ (orange), $\Delta P = 1$ (green),
		and $\Delta P = 2$ (blue).}
		\label{FigDmu05}
	\end{figure}

\begin{figure}
		\begin{center}
			\includegraphics[scale=0.5]{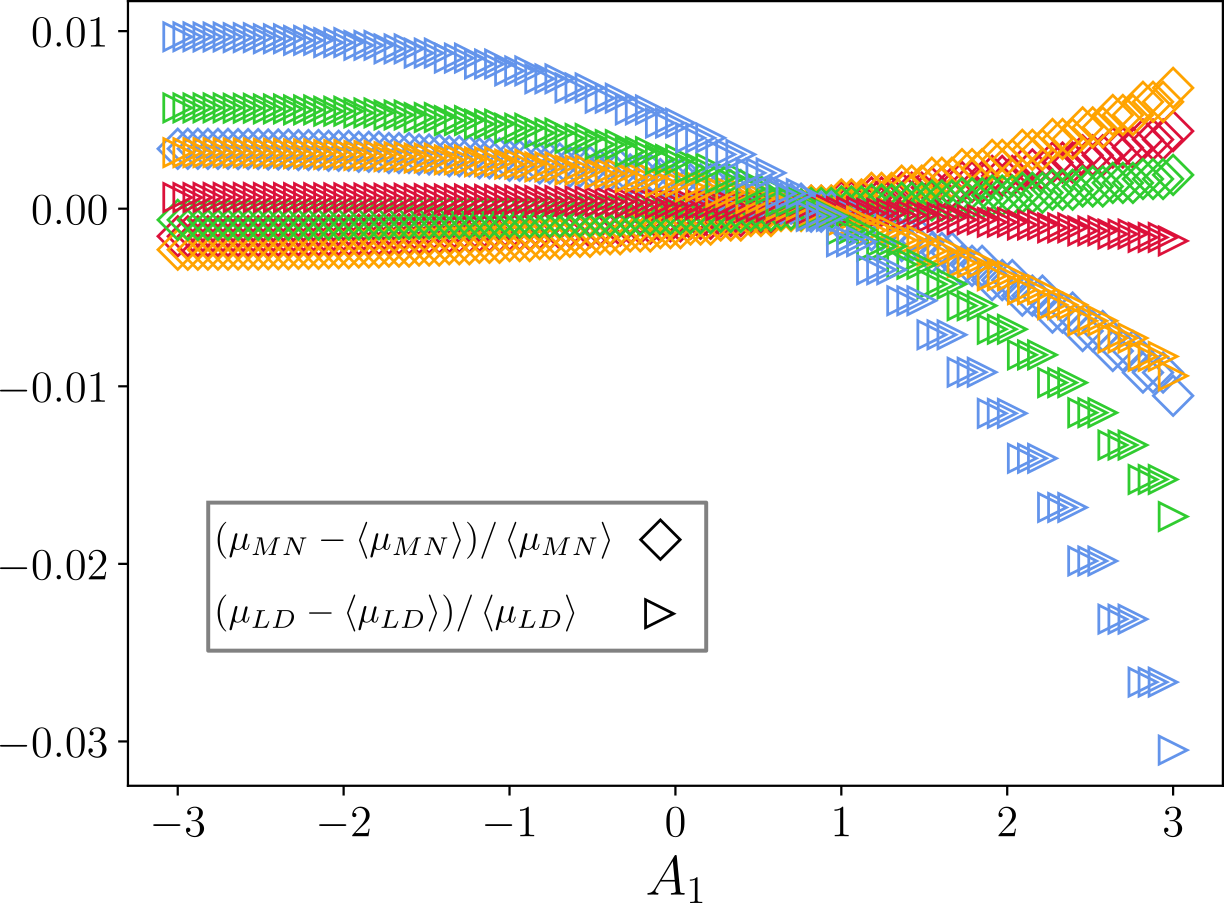}
		\end{center}
		\caption{Relative variation of the Matsuoka-Nakai (diamonds) and Lade-Duncan (triangles) effective friction coefficients for various values of $\Delta P$.
		The parameters are $S_0=S_1=1$, $P_0=1$, $\gamma_c=0.4$, $\Delta P = 0.05$ (red), $\Delta P = 0.5$ (orange), $\Delta P = 1$ (green),
		and $\Delta P = 2$ (blue).}
		\label{FigDmu1}
	\end{figure}

\begin{figure}
		\begin{center}
			\includegraphics[scale=0.5]{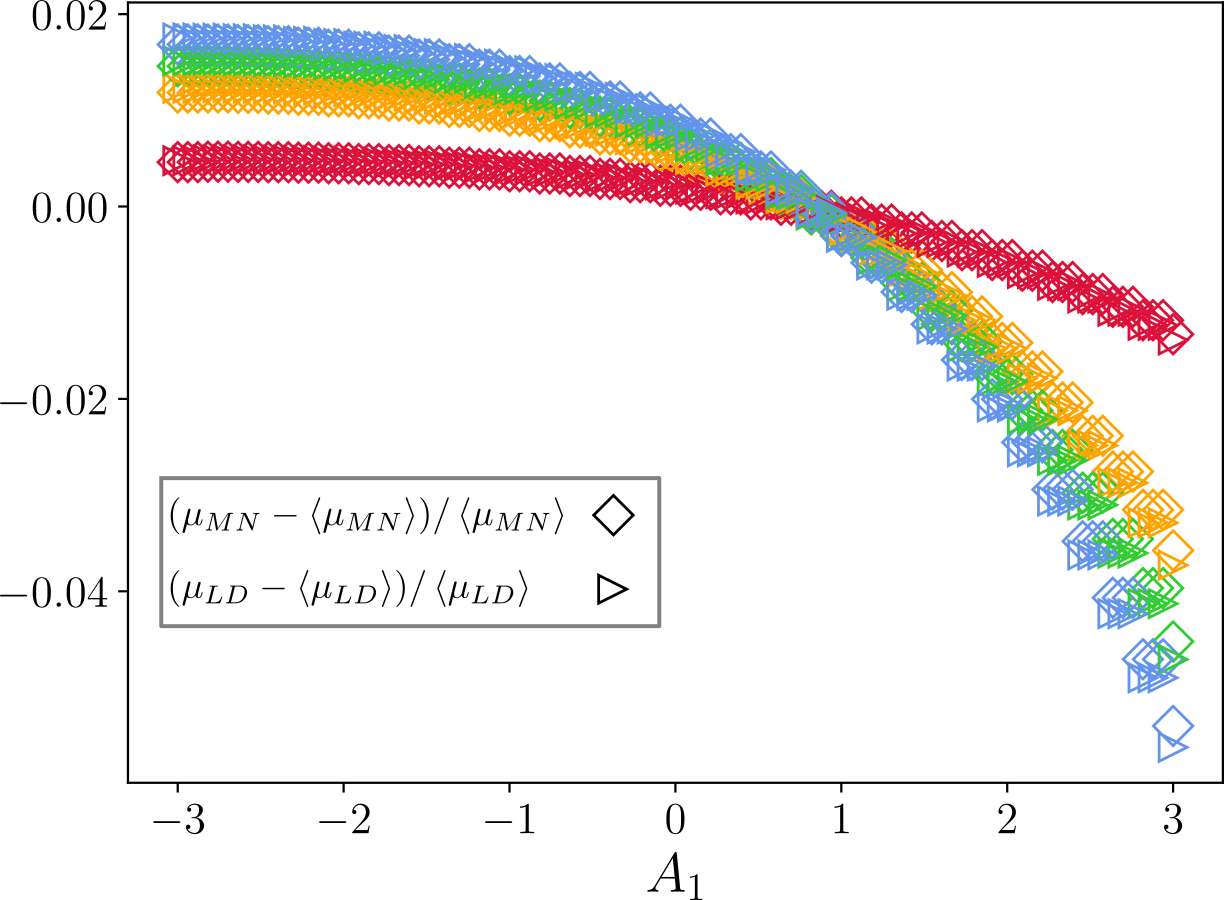}
		\end{center}
		\caption{Relative variation of the Matsuoka-Nakai (diamonds) and Lade-Duncan (triangles) effective friction coefficients for various values of $\Delta P$.
		The parameters are $S_0=S_1=1$, $P_0=10$, $\gamma_c=0.4$, $\Delta P = 0.05$ (red), $\Delta P = 0.5$ (orange), $\Delta P = 1$ (green),
		and $\Delta P = 2$ (blue).}
		\label{FigDmu10}
	\end{figure}

		Then, we can study the variation of the effective friction coefficients as prescribed by the Matsuoka-Nakai and Lade-Duncan criteria, and corresponding
		to definition 2 of the effective friction coefficient.
		The relations between $\mu_{MN}$ (respectively $\mu_{LD}$) and $C_{MN}$ (respectively $C_{LD}$) can be found in \cite{Fleischmann20}.
		Doing so requires to choose a value of $P_0$.
		Results for a sample of values are shown on Figs.~\ref{FigDmu05}, \ref{FigDmu1} and \ref{FigDmu10}.
		It should be kept in mind that comparing the precision to that of the toy-model of this figure makes no sense since Fig.~\ref{FigDDmu} presents
		only the reduced friction coefficient $\Delta\mu$ which necessarily varies more than $\mu$.

		As for the case of the effective friction from the toy model, at least for high enough value of $P_0$, the effective friction is not constant
		in a given deviatoric plane, and even more so that $\Delta P$ is large (or equivalently that the yield surface cut's shape becomes more and more triangular).
		At high values of $P_0$, both effective friction coefficient variations collapse onto each other.
		Perhaps more surprising, it seems to be an optimal value of mild $P_0$ where the variations are reduced at best.

\section{Identification of the strain tensor}

	In order to get a deeper understanding of the physical content of our toy model, let us push it a bit further and replace the screening factor with a
	Heaviside function $\Theta(1 - \dot\gamma t/\gamma_c)$.
	Moreover, since we mainly study the high Weissenberg number regime Wi$\gg1$, the decay of the structural relaxation term $\exp(-\Gamma t)$ happens outside
	of the Heaviside window, our dynamical ansatz reduces to $\Phi_{q(t)}(t) \simeq \Theta(1 - \dot\gamma t/\gamma_c)$.

	Then, using Eq.~(\ref{eqAB3}) and the results of the corresponding appendix, and the fact that the flow is incompressible, 
	the correction to the $j$th component of the stress tensor is given by:
	\begin{equation}
	\label{eqHook}
		\begin{split}
			\Delta \sigma_{jj} &\simeq 2 S_1 \overline{\kappa}_{jj}\int_0^{+\infty}\dot\gamma dt\,\exp\big(2\overline{\kappa}_{jj}\dot\gamma t\big)
			\Theta\big(1 - \dot\gamma t/\gamma_c \big)^2 \\
			& = S_1 \, \Big[\exp(2\overline{\kappa}_{jj}\gamma_c) - 1\Big]\,.
		\end{split}
	\end{equation}
	Let us have a close look at the term inside the brackets.
	The first term has a form of a Finger tensor for a strain given by the typical strain scale $\gamma_c$, and the strength of the flow from which the strain originates,
	$\overline{\kappa}_{jj}$.
	The term inside the brackets can therefore be understood as the difference between the metric in the deformed state, and the unit metric, which is by definition
	the strain tensor (or more precisely here its $j$th component).
	Eq.~(\ref{eqHook}) can thus be straightforwardly read as a Hooke's law, with a shear modulus $S_1$ and a strain tensor $\exp(2\overline{\kappa}_{jj}\gamma_c) - 1$.

	The incompressibility condition Tr$(\kappa)=0$ ensures that the determinant of the deformation tensor is equal to 1, or here that the determinant of the Finger tensor is 1.
	In the limit of small deformations, which corresponds to $\overline{\kappa}_{jj}\gamma_c\ll1$ --- or more precisely $A_2\gamma_c\ll1$, see appendix on the Drucker-Prager limit
	--- one recovers the well known result that $\epsilon$ is fully deviatoric (Tr$(\epsilon) = 0$).
	However, pay attention to the fact that this result only holds in this limit.

	This model is of course too simple to be really accurate, but it shows how $\epsilon$ can be identified in our set of equations.
	When the Heaviside profile is replaced by the Gaussian one used in the paper, the procedure stays the same, although the results are less obvious to interpret since $\epsilon$
	is now expressed as a function of $\mathcal{F}$, which does not let the difference between the strained and unstrained metrics appear explicitly.
	As it turns out, going from the sharp Heaviside profile to the smoother Gaussian one amounts to accounting for the fact that advection is not an instantaneous process,
	and as a result, the final expression of the strain is smoothened.
	The use of a Lorentzian profile, such as the one used in \cite{Brader09} would probably lead to the identification of the same type of function, although in that case,
	the integral cannot be performed exactly.

\bibliography{YP.bib}

\begin{thebibliography}{75}
\bibitem[{foo({\natexlab{a}})}]{foot1}
\bibinfo{note}{Other yield criteria, used primarily to describe the failure of
  metals, such as the Von-Mises' criterion are not discussed here. Indeed, they
  can be directly deduced from the ones we present by neglecting the dependence
  in the pressure invariant $I_1$.}
\bibitem[{foo({\natexlab{b}})}]{foot2}
\bibinfo{note}{Indeed, if the we define an internal friction angle $\alpha$ by
  $\mu_{MC}=\tan(\alpha)$, one can show that the ratio of the length $d_0$
  between the hydrostatic axis $\sigma_1=\sigma_2=\sigma_3$ and the point of
  Lode angle $\theta=0$, and that of Lode angle $\theta=\pi/3$ (both
  represented as big green dots on Fig.~\ref{FigYS}), $d_{60}$, is given by:
  $d_0^2/d_{60}^2=(3+\sin(\alpha))/(3-\sin(\alpha))$. Given that
  $\alpha\in[0;\pi/2]$ by definition, this ratio can be equal to 1 only if
  $\alpha=0$, in which case $\mu=0$ as well.}
\bibitem[{foo({\natexlab{c}})}]{foot3}
\bibinfo{note}{Indeed, there is no obvious expansion parameter in that case.
  Another way of generating polynomial equations is to approximate
  $\mathcal{F}(x)$ by a rational fraction, which gives pretty accurate results
  for the physical region of moderate valies of $x$.}
\bibitem[{foo({\natexlab{d}})}]{foot4}
\bibinfo{note}{The very large $\Delta P$ regime is not explored. Indeed, in
  that case the yield surface has sharper edges which are not so common in the
  experimental literature, so that the relevance of this regime for real,
  physical systems is not obvious.}
\end{thebibliography}

\end{document}